# Silicon Photonic Microring Based Chip-Scale Accelerator for Delayed Feedback Reservoir Computing

Sairam Sri Vatsavai  
Electrical and Computer Engineering  
University of Kentucky  
Lexington, USA  
sairam_srivatsavai@uky.edu  

Ishan Thakkar  
Electrical and Computer Engineering  
University of Kentucky  
Lexington, USA  
igthakkar@uky.edu

*Abstract*— **To perform temporal and sequential machine learning tasks, the use of conventional Recurrent Neural Networks (RNNs) has been dwindling due to the training complexities of RNNs. To this end, accelerators for delayed feedback reservoir computing (DFRC) have attracted attention in lieu of RNNs, due to their simple hardware implementations. A typical implementation of a DFRC accelerator consists of a delay loop and a single nonlinear neuron, together acting as multiple virtual nodes for computing. In prior work, photonic DFRC accelerators have shown an undisputed advantage of fast computation over their electronic counterparts. In this paper, we propose a more energy-efficient chip-scale DFRC accelerator that employs a silicon photonic microring (MR) based nonlinear neuron along with on-chip photonic waveguides-based delayed feedback loop. Our evaluations show that, compared to a well-known photonic DFRC accelerator from prior work, our proposed MR-based DFRC accelerator achieves 35% and 98.7% lower normalized root mean square error (NRMSE), respectively, for the prediction tasks of *NARMA10* and *Santa Fe* time series. In addition, our MR-based DFRC accelerator achieves 58.8% lower symbol error rate (SER) for the Non-Linear Channel Equalization task. Moreover, our MR-based DFRC accelerator has 98× and 93× faster training time, respectively, compared to an electronic and a photonic DFRC accelerators from prior work.**

## I. INTRODUCTION

Artificial Neural Networks (ANNs) have achieved remarkable progress in recent years, and they are being aggressively utilized in real-world applications related to artificial intelligence (AI) and machine learning [1]. In general, ANNs mimic biological neural networks. Depending on the type of the computing task, an ANN architecture can be classified as a feedforward network (FNN) used for static or non-temporal data processing [2], or a recurrent neural network (RNN) that is used for dynamic or temporal data processing [3]. Theoretically, RNNs are very powerful tools for solving complex temporal machine learning tasks. But, the application of RNNs to real world problems is not always feasible due to their high computational training cost and slow convergence [4]. To mitigate these shortcomings, Reservoir Computing (RC) was proposed [5], which is a computational framework [6] derived from the RNN models such as the echo state networks (ESNs) [7] and liquid state machines (LSMs) [8].

In an accelerator for RC (Fig. 1), data inputs are transformed into spatiotemporal patterns in a high dimensional space using a *reservoir*, and analysis of these patterns is performed by an *output* layer. The inputs are connected to the *reservoir* with weights $W_{in}$ at the input layer, and the *reservoir* is connected to the output layer with weights $W_{out}$. The *reservoir* consists of large number of nonlinear (NL) nodes (Fig. 1) that are connected to each other through the recurrent nonlinear dynamics by weights $W_R$. The output of the *reservoir* is the linear combination of $W_{out}$ and the state of the NL nodes connected to the output layer. The key trait of RC is that the input weights $W_{in}$ and the reservoir weights $W_R$ are not trained; they are fixed and random. The output weights $W_{out}$ are trained using simple learning algorithms, e.g., linear regression, thus, remarkably reducing the computational cost of learning, compared to the standard RNNs [9]. Thus, an accelerator for RC can have benefits of both the fast information processing and low learning cost [10], compared to RNN accelerators.

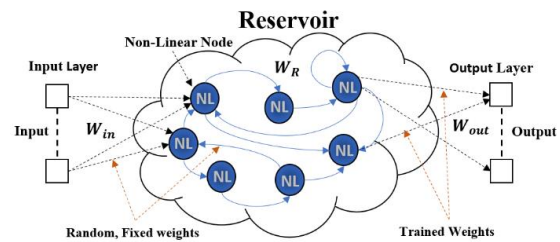

**Fig. 1: Schematic of a reservoir computing (RC) accelerator.**

Photonics based implementations of RC accelerators are attractive as photonics can have low dynamic power consumption and extremely fast computation. Several photonics based RC accelerators have been proposed in the past (e.g., [11]-[18]). However, these implementations do not scale well as they may require up to $10^2$-$10^3$ NL nodes per reservoir [44], making their feasibility a big challenge. In contrast, an alternative model for RC called delayed feedback reservoir computing (DFRC) has been proposed in [19], which employs a dynamic system consisting of a single NL node subjected to delayed feedback [19]. It is shown in prior work (e.g., [26]-[29]) that DFRC accelerators can achieve comparable performance with low hardware overhead compared to the traditional RC accelerators, thereby increasing the ease of implementation and scalability. In fact, prior work also present several photonic DFRC accelerators (e.g., [20]-[23]). But unfortunately, the DFRC accelerators from prior work still require long training times and significantly large area, which limits their applicability to intra-datacenter reservoir computing only. In contrast, *to meet the growing demand of implementing RC-based AI on the edge devices (e.g., for applications related to ubiquitous robotics and smart manufacturing) (Section II.B), realizing a compact DFRC accelerator that can be fully integrated on a chip is of paramount importance*.

In this paper, we present an architecture of a chip-scale photonic DFRC accelerator, which employs a CMOS-compatible active silicon microring resonator (MR) (e.g., [40]) as the NL reservoir node and a low-loss on-chip photonic waveguide (e.g., [40]) as the optical feedback loop. Our DFRC accelerator benefits from its MR reservoir node's rich nonlinearity [17][18] to enable ultra-fast, reasonably accurate, and energy-efficient RC.

[1]This work was supported by a seed grant from the University of Kentucky.

Our contributions in this paper are summarized below:

- We present a compact architecture of an integrated photonic DFRC accelerator that has an active MR as its NL node and a low-loss photonic waveguide as its feedback loop (Section IV);
- We evaluate our MR-based DFRC accelerator's efficiency for performing typical RC tasks such as NARMA10 [32], Santa Fe Time Series [32], and Nonlinear Channel Equalization [5];
- We evaluate our MR-based DFRC accelerator in terms of training time and prediction error metrics, and compare it with a photonic [20] and an electronic [19] DFRC accelerators from prior work (Section V).

## II. RELATED WORK AND MOTIVATION

Several accelerators for RC have been reported that use various types of physical systems, substrates, and devices [11-24], [26-30]. Some of these accelerators (e.g.,[11-12]) use large reservoirs that employ many interconnected NL nodes (as in Fig. 1), whereas the other accelerators (e.g., [20-24], [28-30]) use delayed feedback reservoirs (DFRs) with one physical NL node subjected to delayed feedback to behave as *N* virtual NL nodes (Section III.A). Compared to large reservoirs, DFRs have recently become more popular due to their implementation simplicity. A physical reservoir (DFR or large) needs to satisfy a few traits as discussed in [24], to efficiently solve the temporal computing problems. These traits for DFRs mainly depend on the NL nodes. Prior works on DFRC accelerators explore several types of DFR NL nodes, as discussed next.

### A. DFRC Accelerators from Prior Work

In general, a DFRC accelerator employs a physical NL node and a delayed feedback loop. The DFRC accelerators from prior work broadly use either electronic or photonic implementation for the NL node and delayed feedback loop. The electronic DFRC accelerators (e.g., [10,28-29]) typically use analog circuits to implement NL transformation. These analog circuits also take care of the required delayed feedback. However, these analog circuits can typically suffer from high capacitive loading, which can significantly reduce the speed and power-efficiency, especially when longer delay is required to accommodate large number of virtual nodes. In contrast, photonic DFRC accelerators (e.g., [20-23] enjoy distance-independent and fast computation speed. They employ NL photonic devices, such as Mach-Zehnder Interferometers (MZIs) [21], semiconductor lasers [22], Vertical Cavity Surface Emitting Lasers (VCSEL) [23], as NL nodes. To implement the feedback loop, they typically use a bulky fiber spool [20]. Due to the large size of the photonic NL devices (e.g., a few micro-milli meters for MZIs [43]) and long fiber spools (up to 1.7km [20]), these photonic DFRC accelerators generally yield feedback loop delay ($\tau$) in the micro-milli seconds range, which in turn yields significantly long time (a few tens-hundreds of seconds) for these accelerators to collect the reservoir states for output weights training. To decrease the training time, recent work [30] used a deep learning approach with multiple DFRs and fiber spools. However, the use of bulky fiber spools limits the deployment of such DFRC accelerators to high-end computing systems and datacenters only.

### B. Motivation for MR-based Chip-Scale DFRC Accelerators

Due to the emergence of ubiquitous robotics and smart manufacturing, the demand for RC-based information processing and AI is rapidly growing. To this end, bulky fiber spools based photonic DFRC accelerators find limited pertinence. To address this shortcoming, we present a chip-scale DFRC accelerator that uses an active microring resonator (MR) as the NL reservoir node and a low-loss photonic waveguide as the delay feedback loop. In prior work [17] and [18], rich nonlinearity of MR through-port response has been leveraged to realize RC accelerators. However, these MR-based accelerators employ large reservoirs with large number of MR-based NL nodes, and therefore, they do not scale well for integrated applications. In contrast, our proposed DFRC accelerator uses only one active MR as the single physical NL node that can be scaled to *N* virtual nodes on demand using a photonic waveguide based delayed feedback loop, which makes our DFRC accelerator highly viable for chip-scale applications.

## III. BACKGROUND AND FUNDAMENTALS

### A. Delayed Feedback Reservoir Computing (DFRC)

The basic idea of a delayed feedback reservoir computing (DFRC) accelerator is to have a single physical NL node to behave as *N* virtual NL nodes. Fig. 2 illustrates the functioning of a typical DFRC accelerator, which consists of *(i)* pre-processing (masking) of input signals (Fig. 2(a)), *(ii)* generation of *N* DFR states (Fig. 2(b)), and *(iii)* output generation (Fig. 2 (b)), as discussed next.

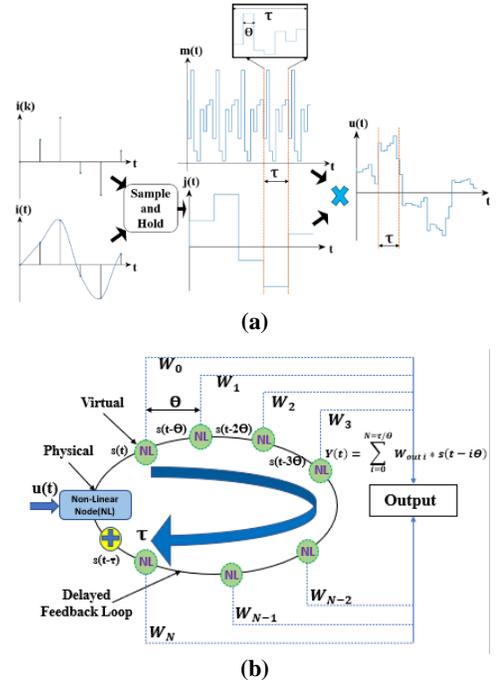

**Fig. 2: Illustration of (a) the pre-processing (masking) of input signal, and (b) the generation of delayed feedback reservoir (DFR) states and final output, for a typical DFRC accelerator.**

#### 1) Pre-Processing (Masking) of Input Signal

A time-dependent input to the DFRC accelerator can be a continuous-time signal $u(t)$ or a discrete-time signal $u(k)$ (Fig. 2(a)). Regardless, such input signal is typically sampled and held to produce a sampled (discretized) continuous signal $j(t)$, with each sample of $j(t)$ being constant for a period of $\tau$. Then, $j(t)$ is multiplied with a periodic masking signal $m(t)$, which plays the role of assigning weights (input weights $W_{in}$; Fig. 1) to the virtual nodes. The masking signal $m(t)$, with its period being $\tau$, varies its value at each $\Theta$ interval for total *N* times during every $\tau$ period (Fig. 2(a) inset), so that $\tau = N*\Theta$. The masking is essential for sequentializing the input, breaking the symmetry of the input and enabling the high dimensional space [24]. The periodic nature of $m(t)$ (i.e., $m(t)$ holds the same value for a corresponding $\Theta$ in every $\tau$ period) ensures that the weight assigned to each virtual node remains identical for all input samples. Multiplying *m(t)* with *j(t)* generates the masked signal *u(t)* that is given as input to the DFR.

*2) Generation of Delayed Feedback Reservoir (DFR) states*

From Fig. 2(b), the physical NL node of the delayed feedback reservoir (DFR) nonlinearly transforms *u(t)* to generate the state of the DFR *s(t)*, and then propagates *s(t)* along the delayed feedback loop. The state of the DFR is generated at each *Θ* interval, hence producing total *N* state values (i.e., *s(t), s(t- Θ), s(t-2Θ), …, s(t-NΘ)*) in each *τ* period. Each of these *N* state values represents the output of the corresponding virtual NL node (total *N* virtual nodes). The input to the physical NL node is the combination of *u(t)* for current *τ* period and the DFR response *s(t- τ)* (from the feedback loop) that was obtained for *u(t-τ)* of the previous *τ* period. The governing formula for the DFR states is given in Eq. (1-2) [19].

$$s(t - i\theta) = FNL(s(t - \tau), u(t), \theta) \quad 0 \leq i \leq N \quad (1)$$

$$u(t) = j(t) * m(t) \quad (2)$$

where *FNL* is the nonlinear transfer function of the physical NL node. The time separation between the two virtual NL nodes is *Θ*. *N* is the number of virtual nodes and *τ* corresponds to the total delay of the feedback loop.

*3) Training of Output Weights for Final Output Generation*

The states of the virtual NL nodes of the DFR are connected to the output layer with the output weights $W_{out,0}$, $W_{out,1}$, …, $W_{out,N}$ (Fig. 2(b)). The final output *Y(t)* of the DFR is the linear combination of these output weights and the states of the DFR, as given by Eq. (3) [19].

$$Y(t) = \sum_{i=0}^{N=\tau/\theta} W_{out,i} * s(t - i\theta) \quad (3)$$

where $W_{out,i}$ are the output weights, and *s(t-iΘ)* are the DFR states. The output weights of the DFR are typically trained offline using lightweight algorithms (e.g., linear regression). During training, the output weights are initialized, then the states of the DFR are observed for the input, and the corresponding intermediate output of the DFR is calculated. The deviation in the intermediate DFR output from the correct output (known during training) is referred to as training error. Then, the output weights are optimized using a learning algorithm until the least value of the training error is achieved. Several techniques, such as Least Mean Squares, Recursive Least Mean Squares and Moore-Penrose Pseudo Inverse [24], can be used to converge to optimal output weights during training. In this paper, we use the Moore-Penrose Pseudo Inverse technique as it suffers less from converging to local minima.

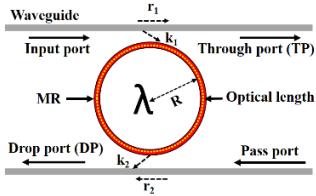

**Fig. 3: An add-drop microring resonator (MR) with radius R, resonance wavelength λ, and coupling waveguides with cross-coupling coefficients $k_1$, $k_2$ and self-coupling coefficients $r_1$, $r_2$ [40].**

*B. Fundamentals: Microring Resonators (MRs)*

A microring resonator (MR) is an optical waveguide looped back on itself as shown in Fig. 3. It also consists of a coupling mechanism to access the loop. The MR is in resonance when the optical path length inside the MR cavity is an integer multiple of the input wavelength. The incident optical at the input port of the MR is transmitted to the through port and to the drop port (Fig. 3). From [18], the drop-port transmission of an active MR shows very rich power-dependent nonlinear response, due to the two-photon absorption (TPA) effects [18]. This rich nonlinearity of MR's through-port response has been leveraged in [17] and [18] to realize efficient RC accelerators. In this paper, we propose to use this rich nonlinearity to implement the physical NL node of our DFRC accelerator, as discussed next.

IV. PROPSOED MR-BASED CHIP-SCALE DFRC ACCELERATOR

*A. Overview*

Fig. 4 illustrates our proposed MR-based DFRC accelerator with photonic waveguide as a feedback loop. The accelerator architecture can be divided into three layers: *(i)* input layer, *(ii)* reservoir layer, and *(iii)* output layer. The input layer consists of a laser source (off-chip) that injects continuous-wave (CW) light of wavelength λ into the on-chip waveguide. The on-chip waveguide is coupled to an MR modulator that modulates the input CW λ with respect to the masked input *u(t)* to generate a modulated optical signal (optical *u(t)*) in the waveguide at the through port of the MR modulator. From [18], the through port response of an MR modulator does not show significant nonlinearity, therefore, optical *u(t)* at the through port of the MR modulator in the input layer is not nonlinearly transformed. This optical *u(t)* propagates along the waveguide and enters the reservoir layer.

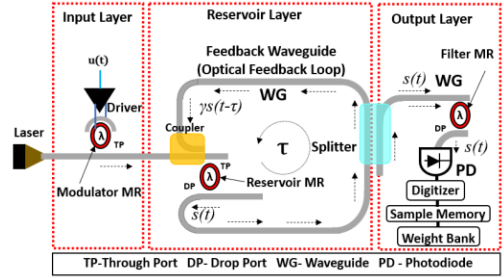

**Fig. 4: Schematic layout of our proposed MR-based DFRC accelerator. The parts enclosed in the red colored boxes can be integrated on a chip. PD = photodiode.**

In the reservoir layer, the input waveguide is coupled with one end of the feedback waveguide (used for delayed feedback loop) via a coupler. This coupler sums the input optical signal *u(t)* with the feedback optical signal *γs(t-τ)*, provided from the feedback waveguide that imposes the propagation delay of *τ*. Just like *u(t)* (Section III.A), the feedback optical signal *γs(t-τ)* also changes every *Θ* interval, and therefore, the summed optical signal (*u(t)+γs(t-τ)*) at the coupler also changes every *Θ* interval. This (*u(t)+γs(t-τ)*) signal then passes through the reservoir MR that acts as the NL node in the reservoir. The drop port of the reservoir MR is coupled with the other end of the feedback waveguide. At the drop port of the reservoir MR (NL node), the (*u(t)+γs(t-τ)*) signal is nonlinearly transformed into *s(t)*, which is then injected into the feedback waveguide. Just like (*u(t)+γs(t-τ)*), *s(t)* also changes every *Θ* interval during the feedback period *τ*. The value of *s(t)* during every *Θ* interval corresponds to the state of a virtual NL node. Thus, to have *N* virtual NL nodes in the system, total *N* *Θ*-intervals should be accomodated in period *τ*. The reservoir signal *s(t)* travels in the feedback waveguide for total delay of *τ* and experiences attenuation by factor *γ* due to optical losses. The resultant signal

$\gamma s(t)$ combines with $u(t+\tau)$ at the input to the reservoir MR, for the next $\tau$ step. Eq. (6-7) in Section IV.B show how the reservoir MR (NL node) transforms $(u(t)+\gamma s(t-\tau))$ into $s(t)$.

The output layer of the DFRC accelerator is connected to the reservoir layer with a splitter, which splits a fractional power of the optical signal $s(t)$ travelling in the feedback waveguide and transfers it to the output layer. At the output layer, we employ another MR that acts as a filter to sample $s(t)$ at each $\Theta$. Total $N$ such samples obtained during $\tau$ period represents the state of the reservoir for $\tau$ period. These $N$ samples are converted by a photodiode (PD) into electrical domain, and are then digitized by the digitizer before being stored in the sample memory. These digitized samples in the sample memory are used for training the output weights that can be stored in the weight bank. During the testing/operating phase, the trained weights from the weight bank can be used to transform the sampled reservoir state to predict the output of the task.

*B. Modelling the Nonlinear Response of the Reservoir MR*

From [35], the richly nonlinear response (due to TPA) of an active MR's through-port transmission depends on the photon lifetime ($\tau_{ph}$) of the MR cavity. For an MR, $\tau_{ph}$ depends on the MR's Q-factor. Therefore, we propose to control the nonlinearity of the reservoir MR (NL node), by controlling the Q-factor (hence, $\tau_p$) of the MR. We can vary the Q-factor (hence, $\tau_{ph}$) by doping the MR with a PN-junction and applying a reverse bias voltage across it to change the Q-factor (hence, $\tau_{ph}$) [42]. We model the $\tau_{ph}$ dependent NL through-port response of the reservoir MR using Eq. (6-7).

$$s(t) = (u(t) + \gamma.s(t-\tau)).\left(1 - e^{(-\theta/\tau_{ph})}\right) \\ + s(t-\tau), \quad if \quad u(t) > s(t-\theta) \quad (6)$$

$$s(t) = (u(t) + \gamma.s(t-\tau)).\left(1 - e^{(-\theta/\tau_{ph})}\right) \\ + s(t-\tau).\left(e^{(-\theta/\tau_{ph})}\right), \quad if \quad u(t) < s(t-\theta) \quad (7)$$

Definitions of the relevant terms in these equations are given in the text of Section IV.A.

## V. EVALUATION

*A. Evaluation Setup*

We evaluate the performance of our proposed MR-based DFRC accelerator for three RC tasks that include two time series prediction tasks such as NARMA10 [32] and Santa Fe [33], and the third Nonlinear Channel Equalization task [5]. The timeseries prediction tasks have important application, both in engineering and medical care [31]. The computational abilities of the proposed DFRC accelerator are examined using the Normalized Root Mean Square Error (NRMSE) and Symbol Error Rate (SER). Our proposed MR-based DFRC accelerator (henceforth, identified as 'Silicon MR') is compared with the Mackey-Glass (MG) differential delay model based electronic DFRC accelerator [19] (henceforth, identified as 'Electronic (MG)') and the MZI-based all optical DFRC accelerator [20] (identified as 'All Optical (MZI)'). In our setup, we use the binary masking technique proposed in [25], which uses maximum length sequences (MLS) to generate optimal mask pattern that is suitable across various tasks. To ensure ideal comparison with prior works, we employ the same binary masking technique across all considered DFRC accelerators. We analyzed the prediction errors and training time for 'Electronic (MG)', 'All Optical (MZI)' and 'Silicon MR' accelerators across the considered benchmark tasks. We also report and discuss the total power consumption for our considered photonic DFRC accelerators.

*B. Error Metrics*

Each considered benchmark task has a dataset, which consists of input signals and corresponding target output. Generally, the dataset is divided into two subsets called training and test sets. The training set is used to train the output weights of the DFR with the goal to predict the target output. The deviation in the target output from the predicted output is called error. The error reported for the DFRC accelerator on the test set gives the performance; lower the error better the DFRC accelerator performance. Below are the error metrics we have used in this paper.

*1) Normalized Root Mean Square Error(NRMSE)*

We use NRMSE for NARMA10 and Santa Fe timeseries tasks. NRMSE is defined by Eq. (8) [20].

$$NRMSE = \sqrt{\frac{\sum_{i=1}^{N}(y_i - \hat{y}_i)^2}{N\sigma_{\hat{y}}^2}} \quad (8)$$

where $\hat{y}$ is the predicted output, $y_i$ is the target output, $N$ is the total length of test set, $\sigma^2_{\hat{y}}$ is the variance of target output.

*2) Symbol Error Rate (SER)*

SER is used only for the Nonlinear Channel Equalization task, in which the DFR reproduces the input symbol in a noisy channel. The SER is evaluated using Eq. (9) [5].

$$SER = \frac{Total\ number\ of\ correctly\ reproduced\ symbols}{Total\ Number\ of\ symbols} \quad (9)$$

*C. Prediction Error Evaluation*

*1) NARMA10*

The Nonlinear Autoregressive Moving Average of 10th order (NARMA10) is a time series whose current output depends on the past ten outputs. It was introduction in [37]. NARMA10 is a standard benchmark task for RC with DFRC accelerators [19-20]. For NARMA10, the input $i(k)$ is drawn from a uniform distribution in interval [0, 0.5] and the target output $y(k+1)$ is given by Eq. (10).

$$y(k+1) = 0.3y(k) + 0.05y(k)\left[\sum_{i=0}^{9}y(k-i)\right] \\ + 1.5i(k)i(k-9) + 0.1 \quad (10)$$

The input $i(k)$ is converted to $j(t)$ by sample and hold, then the masking signal $m(t)$ generated by MLS technique [25] is multiplied with $j(t)$ to get $u(t)$. We generate NARMA10 series dataset with 2000 samples; 1000 samples for training and next 1000 samples for testing, as done in [20]. We report the NRMSE on the testing set. The masking signal, training set and testing set are kept constant across all three considered DFRC accelerators. The NRMSE values for 'Electronic (MG)', 'All Optical (MZI)' and 'Silicon MR' accelerators are shown in Fig. 5. The value of NRMSE depends on the number of virtual nodes ($N$), and therefore, we do a sensitivity analysis to find the optimal value of $N$ for each DFRC accelerator to get the least possible NRMSE. Owing to the space constraints we do not report all results obtained from the sensitivity analysis. In addition, for 'Silicon MR', photonic lifetime ($\tau_{ph}$) (Eq. (6-7)) also directly affects NRMSE. We find the minimum NRMSE for 'Silicon MR' at $N = 900$ and $\tau_{ph} = 50$ ps. Similarly, minimum NRMSE is achieved for 'All Optical (MZI)' and 'Electronic (MG)' at $N = 400$ and 900, respectively. Moreover, 'Silicon MR' achieves 35% lower NRMSE compared to 'All Optical (MZI)', and it performs on par with 'Electronic (MG)', with 'Silicon MR' having very high training speed as discussed in Section V.D. These results clearly corroborate the capabilities of 'Silicon MR' to perform complex RC tasks with good performance.

*2) Santa Fe Time Series*

The Santa Fe timeseries was introduced in [33], which has various datasets labelled A to F. Santa Fe dataset-A is a widely used for evaluating DFRC accelerators [30]. Santa Fe dataset-A records a far-infrared laser operating in chaotic state values. The goal is to predict the laser behavior one-time step ahead. We have used an extended version of Santa Fe dataset-A with 6000 samples from [39]. The dataset is split into 4000 training samples and 2000 testing samples. We obtain the lowest NRMSE for 'Silicon MR' at N=40 and $\tau_{ph}$=50ps. As illustrated in Fig. 5, 'Silicon MR' performs extremely well on Santa Fe compared to 'All Optical (MZI)' with 98.7% lower NRMSE. Even though 'Electronic (MG)' achieves a little better NRMSE, it requires N= 400, which is 10× the N required by 'Silicon MR'.

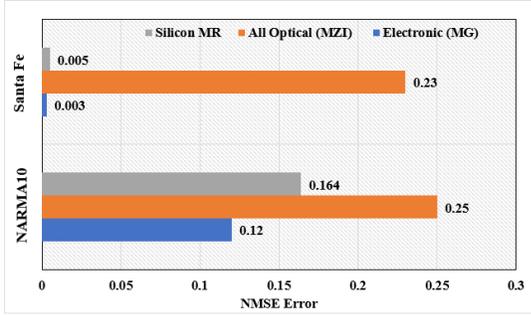

**Fig. 5:** NRMSE values for 'Silicon MR', 'All Optical (MZI)' and 'Electronic (MG)' for NARMA10 and Santa Fe timeseries tasks.

*3) Nonlinear Channel Equalization Task*

In this task, the goal of the DFRC accelerator is to reproduce the distorted symbol due to noise in a wireless communication channel. It is a common task which investigates the nonlinearity of DFRC accelerators. The input-output relation is defined by Eq. (11-12) [5].

$$\begin{aligned}q(n) &= 0.08d(n+2) - 0.12d(n+1) + d(n) \\ &+ 0.18d(n-1) - 0.1d(n-2) \\ &+ 0.09d(n-3) - 0.05d(n-4) \\ &+ 0.04d(n-5) + 0.03d(n-6) \\ &+ 0.01d(n-7)\end{aligned} \quad (11)$$

$$x(n) = q(n) + 0.036q^2(n) - 0.011q^3(n) + v(n) \quad (12)$$

where $d(n)$, is an independent, identically distributed, four level {-3, -1, 1, 3} sequence, $v(n)$ is a pseudo-random Gaussian sequence with zero mean and variance determined by the desired output signal-to-noise ratio (SNR). We vary SNR from 12dB–32dB with a step size of 4dB. We generate 9000 symbols for a dataset, of which 6000 are for training and 3000 are for testing. To evaluate DFRC accelerators for Nonlinear Channel Equalization, we use SER metric. The best SER for 'Silicon MR' is found at $N$=30 and $\tau_{ph}$=50ps. The SERs of 'Silicon MR', 'Electronic (MG)' and 'All Optical (MZI)' are presented in Fig. 6. Across almost all SNR values, the best SER is achieved for 'Electronic (MG)', closely followed by 'Silicon MR' and highest SER for 'All Optical (MZI)'. Nevertheless, 'Silicon MR' reaches 23% lower SER than 'Electronic (MG)' for 24dB SNR. On average, 'Silicon MR' outperforms 'All Optical (MZI)', having 58.8% lower SER.

*D. Training Time*

Fig. 7 gives the time consumed by each DFRC accelerator to complete the training of the output weights. The training time includes time required to observe the DFR states for each input and time required to train the output weights using linear regression. The time involved in generating the DFR states depends on the total delay τ associated with the feedback loop. 'All Optical (MZI)' and 'Silicon MR' clearly have the advantage of shorter training time due to their shorter τ values of 7.56µs and 45ns, respectively, compared to τ = 10ms for 'Electronic (MG)'. Furthermore, the integrated dynamics of 'Silicon MR' yield $\tau_{ph}$ and $\Theta$ to operate at picosecond scale, resulting in a significant speedup of training for 'Silicon MR' by 98× and 93× on average compared to 'All Optical (MZI)' and 'Electronic (MG)' respectively.

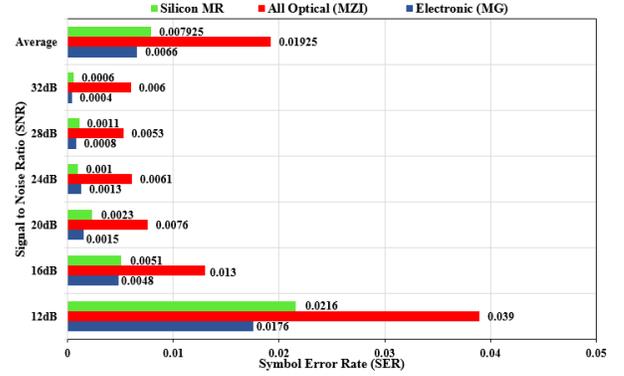

**Fig. 6:** SER values of 'Silicon MR', 'All Optical (MZI)' and 'Electronic (MG)' for Nonlinear Channel Equalization task, with SNR ranging from 12dB-32dB.

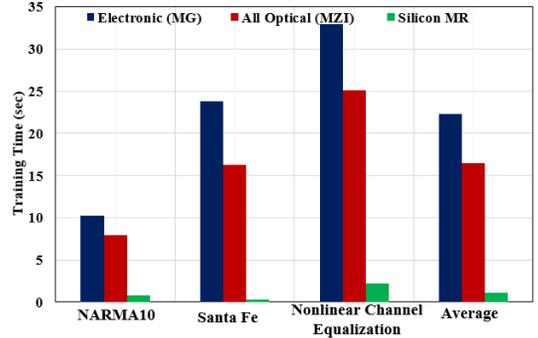

**Fig. 7:** Training time of 'Silicon MR', 'All Optical (MZI)' and 'Electronic (MG)' for tasks NARMA10, Santa Fe and Nonlinear Channel Equalization.

*E. Discussion on Power Consumption*

We report and discuss power consumption of the photonic DFRC accelerators 'Silicon MR' and 'All Optical (MZI)'. In contrast, as the 'Electronic (MG)' implementation in [19] used all computer-controlled off-the-shelf electronic components, we are not able to report the exact power numbers for it. The necessity of ADC and DAC in the reservoir layer of 'Electronic (MG)' hinders its throughput scaling and increases the power consumption compared to photonic DFRC accelerators.

Since the operations of the input and output layers are identical in both photonic DFRC accelerators, major power variations can be noticed in the reservoir layer. For the photonic reservoir in 'Silicon MR' and 'All Optical (MZI)', we calculate the required laser power $P_{Laser}$ using Eq. (15).

$$P_{Laser} = IL^{dB} + \text{Coupling Loss} + \text{Splitter Loss} + \text{Dynamic Range} + S \quad (15)$$

where $IL^{dB}$ is the total insertion loss, dynamic range is optical power range required to implement masking of optical signal, and S is the sensitivity of the photodetector. We considered the values listed in Table 1, and evaluated the total power consumption (laser + dynamic for all layers) in 'Silicon MR' to be 126.48mW and in 'All Optical (MZI)' to be 549.54mW. Due to the high optical resolution of the reservoir MR used in 'Silicon MR', compared to the MZI modulator used in 'All Optical (MZI)', 'Silicon MR' requires lower dynamic range (Table 1) to implement the masking of the input optical signal, thereby, requiring total overall power.

TABLE 1. VARIOUS LOSS AND POWER PARAMETERS

| Parameter | 'Silicon MR' | 'All Optical (MZI)' |
|---|---|---|
| Laser wall-plug efficiency | 10% [35] | 10% [35] |
| PD Sensitivity at 10Gb/s | -5.8dBm [37] | -5.8dBm [37] |
| $IL^{dB}$ | 8.25dB [36] | 7.4 dB [20] |
| Splitter Loss | 0.5dB [38] | NA |
| Coupling Loss | 2dB [38] | 3.3 dB [36] |
| Free Spectral Range (FSR) | 20 nm [39] | NA |
| Dynamic Range | 6dB | 20dB [20] |
| ZHL-32A amplifier | NA | 10 dBm [20] |
| Feedback Photodiode (TTI TIA525) | NA | 1.2mW [20] |
| Optical Attenuator (Agilent 81571A) | NA | 33dBm [20] |
| MR Modulator | 15fJ/bit [37] | NA |
| MR filter | 0.705pJ/bit [37] | NA |
| MZI modulator | NA | 100mW [20] |

## VI. CONCLUSION

In this paper, we presented a silicon MR based chip-scale accelerator for delayed feedback reservoir computing (DFRC). Our DFRC accelerator leverages the rich nonlinearity of the active MR to realize the nonlinear node in the reservoir layer of the accelerator. Moreover, it uses a photonic waveguide as the feedback delay loop to enable fully on-chip integration of the reservoir layer. Evaluation with benchmark tasks shows that our MR-based DFRC accelerator achieves 35% and 98.7% lower NRMSE, up to 58.8% less average SER, and up to 93x faster training time compared to a photonic DFRC accelerator from prior work. Thus, our MR-based DFRC accelerator represents an attractive solution for realizing scalable reservoir computing for integrated applications.